\documentclass[amsmath,11pt]{article}

\usepackage{amssymb,amssymb,epsfig,bm}
\usepackage{caption}
\usepackage{subcaption}
\usepackage{xcolor}

\date{\empty}

\begin{document}

\title{\bf The reason peculiar velocities grow faster in general relativity than in Newtonian gravity}\vspace{15pt}
\author{Erick Pasten${}^1$\footnote{erick.contreras@usm.cl} and Christos G. Tsagas${}^{2,3}$\footnote{email:tsagas@astro.auth.gr}\\
{\small ${}^1$Departamento de F\'{i}sica, Universidad de Santiago de Chile}\\ {\small Avenida V\'{i}ctor Jara 3493, Estaci\'{o}n Central, 9170124, Santiago, Chile}\\ {\small ${}^2$ Section of Astrophysics, Astronomy and Mechanics, Department of Physics}\\ {\small Aristotle
University of Thessaloniki, Thessaloniki 54124, Greece}\\ {${}^3$\small Clare Hall, University of Cambridge, Herschel Road, Cambridge CB3 9AL, UK}}

\maketitle

\begin{abstract}
\textit{Context:} We provide a theoretical comparison between the Newtonian and quasi-Newtonian studies and the general relativistic analysis of linear cosmological peculiar velocities, in view of the increasing number of surveys reporting bulk peculiar flows much faster and considerably deeper than typically expected.\\
\textit{Aims:} We look for a theoretical answer to the ongoing bulk-flow question to explain the fast and deep bulk peculiar motions reported by many recent surveys. These reports have come to be treated as a potentially serious problem for the Lambda cold dark matter ($\Lambda$CDM) model since they claim peculiar velocities well in excess of those permitted by the current cosmological paradigm.\\
\textit{Methods:} We studied the growth of linear peculiar velocities by employing a general relativistic cosmological perturbations theory and then compared our analytical results with those adopted by the $\Lambda$CDM paradigm, which are Newtonian in nature. We did so by means of a unified analysis that facilitates the direct and unambiguous comparison between the Newtonian and the relativistic approaches. This allowed us to identify the differences between the two approaches and to unravel the physical precesses responsible for them.\\
\textit{Results:} Our analysis demonstrates that general relativity leads to a linear growth rate of $v\propto t$, at a minimum, for the peculiar-velocity field ($v$) during the Einstein-de Sitter epoch of the universe, namely between recombination and the onset of the recent phase of accelerated expansion. The Newtonian and quasi-Newtonian studies, on the other hand, have led to the substantially weaker growth rate of $v\propto t^{1/3}$ during the same period. The reason behind the difference is that the relativistic analysis also accounts for the gravitational input of the `peculiar flux', that is of the momentum density triggered by the peculiar motion of the matter. We recall that peculiar flows are nothing but matter in motion and that moving matter means non-zero energy flux. We also recall that in general relativity, as opposed to Newtonian gravity, energy fluxes gravitate because they too contribute to the energy-momentum tensor of the matter fields. We show that this extra input of the peculiar flux drastically modifies the driving force of the peculiar-velocity field and, in so doing, it increases the linear velocity growth well beyond the current (Newtonian-based) $\Lambda$CDM expectations.\\
\textit{Conclusions:} We show that general relativity supports substantially stronger growth rates for the linear peculiar velocities compared to the Newtonian and quasi-Newtonian studies, which in turn leads to faster and deeper residual bulk peculiar flows today. Therefore, one could solve the ongoing bulk-flow puzzle without breaking away from the general framework of the $\Lambda$CDM paradigm by simply employing Einstein's theory instead of Newtonian gravity.
\end{abstract}

\section{Introduction}\label{sI}
Large-scale peculiar motions, commonly known as bulk flows, are an observational certainty confirmed by numerous surveys over several decades. These surveys largely agree on the direction of the reported bulk flows, but they seem to disagree on their size and speed. Moreover, while many of the reported bulk flows have sizes and velocities within the theoretical boundaries of the current cosmological model (e.g.~see~\cite{ND}-\cite{Qetal}), there has been an increasing number of recent surveys reporting bulk peculiar flows that are considerably faster and deeper than theoretically expected (e.g.~see~\cite{WFH}-\cite{WHD}). To these claims, one could also add the controversial `dark flows', with sizes and velocities well in excess of the Lambda cold dark matter ($\Lambda$CDM) limits~\cite{Ketal,KA-BE}. Typically, the surveys that agree with the $\Lambda$CDM expectations cover scales smaller than 100~Mpc, whereas those that disagree  extend beyond the aforementioned threshold. Currently, the reason for this discrepancy remains unknown and a subject of debate within the cosmological community since, if confirmed, the bulk-flow question could prove a serious problem for the concordance the $\Lambda$CDM model. The reader is referred to~\cite{TPA} for a recent review on cosmological peculiar motions, with more details and references.

When it comes to peculiar motions, the $\Lambda$CDM predictions are built on Newtonian theory, which argues for a relatively mild linear growth rate of $v\propto t^{1/3}$ for the peculiar-velocity field ($v$)~\cite{P1,P2}. This was also the result of the `quasi-Newtonian' treatments~\cite{M,EvEM}, though they reduce to mere Newtonian despite their initially relativistic profile. One of the reasons for remaining within the Newtonian limits was probably the anticipated technical difficulty of addressing the bulk-flow question relativistically. Another reason concerns the widespread expectation that when dealing with non-relativistic peculiar motions, Einstein's theory was unlikely to drastically change the existing Newtonian picture. It is therefore not surprising that fully relativistic linear studies of peculiar velocities have only recently appeared in the literature~\cite{TT}-\cite{T}. These treatments argue for a considerably stronger peculiar growth rate, namely for $v\propto t^{4/3}$, which has the potential of addressing both the higher speeds and the greater depths of the bulk flows reported in~\cite{WFH}-\cite{WHD}, for example. The reason for the profound difference between the results of the Newtonian and quasi-Newtonian studies on the one hand and the relativistic studies on the other, can be tracked down to the fundamentally different way the two theories treat the gravitational field and its sources and more specifically to the entirely different role played by the `peculiar flux' in these two theoretical approaches.

Peculiar flows are nothing but matter in motion, and moving matter means a non-zero energy flux (e.g.~see~\cite{TCM,EMM}). In general relativity, as opposed to Newtonian gravity, energy fluxes contribute to the stress-energy tensor and therefore gravitate~\cite{TT}. The gravitational input of the peculiar flux, which survives at the linear perturbative level, is not accounted for in the Newtonian treatments nor in their quasi-Newtonian counterparts, though for different reasons. In the studies using Newtonian treatments, this happens by default, while in the quasi-Newtonian studies, it is due to the strict constraints the quasi-Newtonian approximation imposes on the perturbed spacetime. These compromise the relativistic nature of the approach and reduce it to mere Newtonian for all practical purposes. Although the problematic nature of the quasi-Newtonian approach has been known (e.g.~see related `warning' comments in Sect.~6.8.2 of~\cite{EMM} - pages: 150 and 151), this has not been widely appreciated. The likely reason is the absence of a direct comparison with a fully relativistic analysis (until very recently; see~\cite{T} for such a comparison).

In a fully relativistic linear study, the gravitational input of the peculiar flux feeds into the associated conservation laws and subsequently into the linear evolution formulae of the peculiar-velocity field. In so doing, the relativistic treatment brings into play the effects of density perturbations, namely of structure formation, in addition to those of the peculiar flux. All this drastically modifies the agent driving the linear peculiar velocities. As a result, unlike its Newtonian and quasi-Newtonian counterparts, the relativistic driving agent no longer decays with time. This in turn ensures that the relativistic differential equations and their associated solutions differ considerably. More specifically, the general relativistic analysis supports a substantially stronger linear growth for peculiar velocities during the Einstein-de Sitter phase of the universe. We demonstrate in this work that even in the `minimalist' scenario where only the gravitational input of the peculiar flux is taken into account, the velocity growth rate is $v\propto t$. Moreover, the latter could potentially increase to the $v\propto t^{4/3}$ rate of~\cite{TT,MiT} when structure-formation effects are also accounted for - a possibility worthy of further investigation. In either case, the resulting velocity field has the potential to explain the speed and the depth of the fast bulk flows reported in~\cite{WFH}-\cite{WHD}. Put another way, general relativity could naturally relax the typical (Newtonian) $\Lambda$CDM limits to accommodate the fast bulk peculiar flows.

This work provides a direct comparison between the Newtonian a,d quasi-Newtonian studies and the relativistic analysis of linear peculiar velocities on a Friedmann-Robertson-Walker (FRW) background with flat spatial sections. The linear nature of all the studies implies that their results formally apply to relatively large scales, typically near and beyond the 100~Mpc threshold, where one feels confident to ignore non-linear effects.

We begin our comparison by setting the theoretical framework of the approaches. We identify the point where they start to diverge from each other and explain why this happens. We then derive the linear evolution laws of the associated peculiar-velocity fields, highlight the differences between the corresponding solutions, and discuss their implications for addressing the ongoing bulk-flow question. In the process, we also highlight the role of the Newtonian acceleration and the relativistic four-acceleration as the driving forces of linear peculiar velocities. More specifically, we show that the stronger the four-acceleration, the faster the driven peculiar-velocity field. Finally, we demonstrate a way of recovering the relativistic results from the Newtonian setup by selectively accounting for certain source-free terms in Poisson's formula that are typically sidestepped. Although, this recovery is rather ad hoc and not uniquely determined, it provides useful analogies between the two types of study. For instance, one could argue that ignoring the above mentioned source-free terms in the Newtonian studies is analogous to bypassing the purely general relativistic gravitational input of the peculiar flux.

We start this work with the traditional study of cosmological linear peculiar velocities, which employs Newtonian gravity and proceeds along the lines of~\cite{P1,P2} and~\cite{Netal} in Sect.~\ref{sNTPVs}. We provide the covariant analogue of the Newtonian analysis in Sect.~\ref{sssNT} to familiarise the reader with the covariant formalism and connect the standard Newtonian analysis of Sect.~\ref{sNTPVs} with the relativistic approach of Sect.~\ref{sssRT}. In the latter section, we also discuss the physical reasons the Newtonian and quasi-Newtonian studies of linear peculiar velocities arrive at considerably weaker results compared to those of the relativistic treatment. In so doing, we highlight the key role played by the four-acceleration, keeping in mind that it is the driving force of the linear peculiar-velocity field (see Sect.~\ref{sssR4-A}). We also outline a way of bridging the aforementioned differences between the Newtonian and the relativistic studies by (selectively) including extra contributions to the Newtonian gravitational potential (see~Sect.~\ref{sEGP}). Finally, we close with a discussion of our results and of their role in the quest for an answer to the bulk-flow puzzle.

\section{Newtonian treatment of peculiar velocities}\label{sNTPVs}
Here, we employ Newtonian theory and adopt co-moving coordinates $\left\{\mathbf{x}\right\}$. These coordinates are related to their corresponding physical counterparts $\left\{\mathbf{r}\right\}$ by means of $\mathbf{x}=\mathbf{r}/a$, where $a=a(t)$ is the cosmological scale factor. As a result, the velocity ($\mathbf{u}$) of a typical observer in the Newtonian version of a perturbed Einstein-de Sitter universe decomposes as
\begin{equation}
\mathbf{u}= H\mathbf{r}+ \mathbf{v}\,,  \label{vu}
\end{equation}
with $H=\dot{a}/a$ representing the Hubble parameter. Also, $\mathbf{v}=a\dot{\mathbf{x}}$, which is the observer's peculiar velocity relative to the universal rest frame, which has typically been identified with the rest-frame of the cosmic microwave background (CMB). Within the adopted cosmological framework, the linear evolution of the peculiar-velocity field is monitored by the associated Euler equation, namely by~\cite{P1,P2}
\begin{equation}
\dot{\mathbf{v}}+ H\mathbf{v}= -{1\over a}\,\mathbf{\nabla}\phi\,,  \label{ldotv}
\end{equation}
where $\phi$ is the peculiar gravitational potential and $\nabla$ is the gradient in co-moving coordinates. The latter acts as a linear source of peculiar-velocity perturbations and satisfies Poisson's formula, that is
\begin{equation}
\mathbf{\nabla}^2\phi= 4\pi Ga^2\rho_0\delta\,,  \label{Poisson}
\end{equation}
with $\delta=(\rho-\rho_0)/\rho_0$ being the familiar density contrast, $\rho$ the matter density, and $\rho_0$ its background value.

Differentiating Eq.~(\ref{ldotv}) in time, recalling that $H=2/3t$ and $\dot{H}=2/3t^2$ to zero order, while keeping up to linear order terms leads to
\begin{equation}
\ddot{\mathbf{v}}+ 2H\dot{\mathbf{v}}- {1\over2}H^2\mathbf{v}= \ddot{\mathbf{v}}+ {4\over3t}\,\dot{\mathbf{v}}- {2\over9t^2}\,\mathbf{v}= -{1\over a}\partial_{t}\left(\mathbf{\nabla}\phi\right)\,.  \label{lddotv1}
\end{equation}
One may ignore the right-hand side of the above equation by assuming that the gradient of the peculiar gravitational potential varies slowly in time, for example. Then, (\ref{lddotv1}) reduces to
\begin{equation}
9t^2{{\rm d}^2\mathbf{v}\over{\rm d}t^2}+ 12t\,{{\rm d}\mathbf{v}\over{\rm d}t}- 2\,\mathbf{v}= 0\,,  \label{lddotv2}
\end{equation}
which accepts the power-law solution
\begin{equation}
\mathbf{v}= \mathcal{C}_1t^{1/3}+ \mathcal{C}_2t^{-2/3}\,,  \label{bfv1}
\end{equation}
on all scales. Therefore, according to Newtonian physics, linear peculiar velocities are expected to grow as $\mathbf{v}\propto t^{1/3}$ after recombination and throughout the Einstein-de Sitter epoch.

One can also refine solution (\ref{bfv1}) by incorporating the source term on the right-hand side of (\ref{lddotv1}). Following~\cite{P1,P2}, we started by recalling that the peculiar gravitational acceleration is
\begin{equation}
\mathbf{g}= -{1\over a}\,\mathbf{\nabla}\phi= Ga\rho_0\int\delta{\mathbf{x}^{\prime}-\mathbf{x}\over ||\mathbf{x}^{\prime}-\mathbf{x}||^3}\,{\rm d}^3\mathbf{x}^{\prime}\,,  \label{Npaccel1}
\end{equation}
Given that $\phi$ and $\delta$ also depend on time, after a relatively straightforward calculation, the temporal derivative of the above leads to the linear time-evolution equation for $\nabla\phi$, namely
\begin{equation}
\left({\nabla}\phi\right)^{\cdot}= -H\mathbf{\nabla}\phi- {3\over2}\,aH^2\mathbf{v}\,.  \label{laux1}
\end{equation}
Note that in deriving this relation we have used the background expressions $\rho_0=3H^2/8\pi G$ and $\dot{\rho}_0=-3H\rho_0$ together with the linear result $\dot{\delta}=-\mathbf{\nabla}\cdot\mathbf{v}$. The latter follows from the continuity equation, which reads $\dot{\rho}=-\mathcal{H}\rho$ at the linear level, with $\rho$ and $\mathcal{H}$ being the perturbed matter density and Hubble parameter, respectively. Setting $\rho=\rho_0(1+\delta)$ and $\mathcal{H}=H+\mathbf{\nabla}\mathbf{v}$, substituting into the linear continuity equation, and keeping up to first-order terms, we obtained the desired relation, namely, at $\dot{\delta}=-\mathbf{\nabla}\cdot\mathbf{v}$. Finally, by combining Eqs. (\ref{ldotv}), (\ref{lddotv1}) and (\ref{laux1}), we arrived at the differential equation
\begin{equation}
\ddot{\mathbf{v}}+ 3H\dot{\mathbf{v}}- H^2\mathbf{v}= \ddot{\mathbf{v}}+ {2\over t}\,\dot{\mathbf{v}}- {4\over9t^2}\,\mathbf{v}= 0  \label{lddotv3}
\end{equation}
and the associated power-law solution~\cite{P1,P2}
\begin{equation}
\mathbf{v}= \mathcal{C}_1t^{1/3}+ \mathcal{C}_2t^{-4/3}\,,  \label{bfv2}
\end{equation}
on all scales. Comparing solutions (\ref{bfv1}) and (\ref{bfv2}) made it clear that the inclusion of the non-zero right-hand side of (\ref{lddotv1}) has no physical and/or practical significance. The rate of the growing mode in the former of the two solutions has remained the same and only the decaying mode has changed.

Our last note is on the peculiar acceleration $\mathbf{g}=-\mathbf{\nabla}\phi/a$ (see Eq.~(\ref{Npaccel1})), which is the driving force of the linear $\mathbf{v}$ field in Newtonian theory. By combining (\ref{ldotv}) with solution (\ref{bfv2}), one immediately realises that the Newtonian force decays with time as
\begin{eqnarray}
\mathbf{g}\propto t^{-2/3}\propto a^{-1}\,,\label{Npaccel2}
\end{eqnarray}
since $a\propto t^{2/3}$, after matter--radiation equality. In Sect.~\ref{sssR4-A} we demonstrate that the same time-decay law also applies to the four-acceleration that drives linear peculiar velocities in the quasi-Newtonian studies. In contrast, the four-acceleration driving the peculiar-velocity field in the relativistic treatments is either constant (at a minimum) or increases with time (see Sect.~\ref{sssR4-A}). This difference offers another way of explaining why the Newtonian/quasi-Newtonian studies lead to the mediocre $\mathbf{v}\propto t^{1/3}$ growth and why general relativity supports faster and deeper bulk flows than generally anticipated.

\section{Covariant treatment of peculiar velocities}\label{sCTPVs}
The reported large-scale peculiar motions are believed to have started as weak peculiar-velocity perturbations that were amplified to the observed bulk flows by structure formation. In this section we provide a `unified' covariant study looking into the linear evolution of peculiar velocities in cosmology. In so doing, we also facilitate direct comparison between the Newtonian, the quasi-Newtonian, and the relativistic treatments of the subject.

\subsection{CMB and bulk-flow frames}\label{ssCMBBFFs}
Studies of peculiar velocities require two relatively moving coordinate systems, with one of them following the peculiar motion of the matter and the other the reference frame of the universe. The latter has been typically identified with the coordinate system of the CMB, namely the frame relative to which we define and measure peculiar motions.

In Newtonian theory, the velocities of the aforementioned observers are related via the familiar Galilean transformation
\begin{equation}
\tilde{u}_{\alpha}= u_{\alpha}+ v_{\alpha}\,,  \label{Galilean}
\end{equation}
where $\tilde{u}_{\alpha}$ and $u_{\alpha}$ are the velocities of bulk-flow observers and of their CMB partners, respectively. The term $v_{\alpha}$ is the peculiar velocity of the former with respect to the latter (with $\alpha=1,2,3$). In relativity, the Galilean transformation is replaced by the (also familiar) Lorentz boost, which in the case of non-relativistic peculiar motions reads
\begin{equation}
\tilde{u}_a= u_a+ v_a\,.  \label{Lorentz}
\end{equation}
Here, in contrast to (\ref{Galilean}), $\tilde{u}_a$ and $u_a$ are the (timelike) 4-velocities of the aforementioned observers, while $v_a$ is the (spacelike) velocity of the peculiar motion (with $a=0,1,2,3$). Put another way, $u_au^a=-1=\tilde{u}_a\tilde{u}^a$ and $u_av^a=0$ by construction. Also, $v^2=v_av^a\ll1$ when dealing with non-relativistic peculiar flows. We note that although the transformations (\ref{Galilean}) and (\ref{Lorentz}) may seem formally identical, they are different because the former involves spacelike vectors only, whereas in the latter the two 4-velocity vectors are timelike.

\subsection{The key role of the peculiar flux}\label{ssKRPF}
The key difference between the Newtonian and the relativistic study of peculiar motions follows from the fact that in relativity, the nature of the matter fields changes between the relatively moving frames and the associated observers. In particular, a cosmic medium that appears perfect to one reference frame will appear imperfect to any other relatively moving coordinate system. The imperfection takes the form of an effective energy flux vector, which is solely triggered by relative-motion effects. If we consider a perturbed Friedmann universe with zero spatial curvature, then when setting the pressure (both the isotropic and the viscous) to zero, we arrive at the linear relations
\begin{equation}
\tilde{\rho}= \rho\,, \hspace{15pt}
\tilde{p}= 0\,, \hspace{15pt} \tilde{\pi}_{ab}= 0  \label{lrels1},
\end{equation}
and
\begin{equation}
\tilde{q}_a= q_a- \rho v_a\,,  \label{lrels2}
\end{equation}
between the CMB and the bulk-flow (i.e.~the tilded) frames. In the above $\rho$, $p$, $\pi_{ab}$, and $q_a$ are respectively the density, the isotropic pressure, the viscosity, and the energy flux of the matter relative to the CMB frame, with their tilded analogues measured in the matter frame. Following (\ref{lrels1}), the density and the pressure (both the isotropic and the viscous) of the matter are the same in the two coordinate systems at the linear level. On the other hand, according to (\ref{lrels2}), there is a difference between the energy fluxes entirely due to relative-motion effects. In fact, Eq.~(\ref{lrels2}) ensures that even if the flux vanishes in one frame, there is a non-zero peculiar flux in all the other relatively moving coordinate systems, entirely due to relative-motion effects. Clearly, one cannot set the flux to zero in both frames simultaneously without switching the peculiar velocity off as well. Put another way, in the presence of peculiar motions, there is always a non-zero energy flux, which in turn means that the cosmic medium is no longer perfect.

When assuming a vanishing flux in the matter frame, we may set $\tilde{q}_a=0$. Then, Eq.~(\ref{lrels2}) guarantees a non-zero peculiar flux equal to $q_a=\rho v_a$ in the CMB frame. The latter survives at the linear perturbative level, and in relativity, it contributes to the perturbed energy-momentum tensor and therefore to the relativistic gravitational field. As a result, the linear conservation laws of the energy and the momentum densities acquire flux-related terms and take the forms
\begin{equation}
\dot{\rho}= -\Theta\rho- {\rm D}^aq_a \hspace{7.5mm} {\rm and} \hspace{7.5mm}  \rho A_a= -\dot{q}_a- 4Hq_a\,,  \label{lcls}
\end{equation}
respectively. In the above equation, $\Theta={\rm D}^au_a$ is the three-divergence of the $u_a$ field (with $\Theta=3H$ in the background), $A_a=\dot{u}_a$ is the associated four-acceleration, and ${\rm D}_a$ is the 3D covariant derivative operator. We note that according to (\ref{lcls}b), the four-acceleration is non-zero, despite the absence of pressure. This is a direct consequence of the fact that in the presence of relative motions, there is always a non-zero energy flux in the system and the cosmic medium can no longer be treated as perfect.\footnote{The linear expression (\ref{lrels2}) guarantees that a non-zero peculiar flux ensures a non-zero four-acceleration as well. The only exception is when $\dot{q}_a=-4Hq_a$, which however corresponds to a `set of measure zero' in probabilistic terms. Moreover, this one and only exception leads to a $v$ field that decays as $v\propto a^{-1}\propto t^{-2/3}$. Since peculiar velocities start weak at recombination, there should be no bulk flows today if they were to decay in time. We recall that even in Newtonian theory, linear peculiar velocities grow as $v\propto t^{1/3}$. See~\cite{P1,P2}.} As it turns out (see Sect.~\ref{ssNQ-NRTs} next), it is the peculiar-flux input to the gravitational field that separates the relativistic studies of peculiar motions from the rest.

\subsection{Newtonian and quasi-Newtonian versus relativistic
treatments}\label{ssNQ-NRTs}
As mentioned above, the observed bulk peculiar flows began as peculiar-velocity perturbations, and they were subsequently amplified by structure formation. The driving agent of the amplification process depends on the theory that one employs to study the evolution of the peculiar-velocity field. In Newtonian physics, the driving force is gravity and, more specifically, the gravitational acceleration. Mathematically, this is reflected in the linear evolution equation of the $v_{\alpha}$ field, which reads
\begin{equation}
\dot{v}_{\alpha}+ Hv_{\alpha}= -\partial_{\alpha}\phi\,,  \label{Ndotv}
\end{equation}
with $\phi$ being the Newtonian gravitational potential, and the gradient is in physical coordinates.\footnote{For the co-moving analogue of expression (\ref{ldotv}), the interested reader is referred to Eq.~(5a) in~\cite{Netal}.} In relativity, gravity is not a force but the manifestation of spacetime curvature. As a result, in a relativistic treatment, linear peculiar velocities are driven by non-gravitational forces, and the evolution of linear peculiar velocities is monitored by
\begin{equation}
\dot{v}_a+ Hv_a= -A_a\,,  \label{GRdotv}
\end{equation}
where $A_a=\dot{u}_a$ is the four-acceleration measured in the reference CMB frame.\footnote{It is straightforward to show that expression (\ref{GRdotv}) also follows from (\ref{lcls}b) after substituting $q_a=\rho v_a$ in the right-hand side of the latter. We also note that in the Newtonian equations, overdots imply convective derivatives (e.g.~$\dot{v}_{\alpha}= \partial_tv_{\alpha}+u^{\beta}\partial_{\beta}v_{\alpha}$), while in their relativistic counterparts, overdots indicate covariant differentiation along the associated 4-velocity field (e.g.~$\dot{v}_a=u^b\nabla_bv_a$).} It is therefore clear that the linear evolution of the $v_a$ field is determined by the form and the impact of the associated driving agent.

\subsubsection{Newtonian treatment}\label{sssNT}
Newtonian peculiar-velocity studies start from the differential equation Eq. (\ref{Ndotv}). Recalling that $\dot{H}=-3H^2/2$ to zero order, the linearised time-derivative of the former leads to
\begin{equation}
\ddot{v}_{\alpha}+ 2H\dot{v}_{\alpha}- {1\over2}\,H^2v_{\alpha}= -\partial_{\alpha}\dot{\phi}\,,  \label{ltv''1}
\end{equation}
since $\partial_t(\partial_{\alpha}\phi)= \partial_{\alpha}\dot{\phi}- H\partial_{\alpha}\phi$ to first approximation. As noted in Sect.~\ref{sNTPVs}, ignoring the right-hand side of the above equation leads to the power-law solution $v=\mathcal{C}_1t^{1/3}+\mathcal{C}_2t^{-2/3}$. However, by appealing to the Jeans' swindle (e.g.~see~\cite{BT}) and writing Poisson's equation as $\partial^2\phi=\rho\delta/2$,\footnote{Hereafter, the gravitational constant $\kappa=8\pi G$ as well as the velocity of light are set to equal unity.} a relatively lengthy but fairly straightforward calculation yields
\begin{equation}
\partial_{\alpha}\dot{\phi}= {1\over2}\,H\partial_{\alpha}\phi+ {3\over2}\,H\dot{v}_{\alpha}\,,  \label{graddotPhi}
\end{equation}
which when substituted into the right-hand side of (\ref{ltv''1}) gives
\begin{equation}
\ddot{v}_{\alpha}+ 3H\dot{v}_{\alpha}- H^2v_{\alpha}= 0\,,  \label{ltddotv2}
\end{equation}
with $H=2/3t$ in the background. The above equation can be solved to give
\begin{equation}
v= \mathcal{C}_1t^{1/3}+ \mathcal{C}_2t^{-4/3}= \mathcal{C}_3a^{1/2}+ \mathcal{C}_4a^{-2}\,  \label{lNv2}
\end{equation}
on all scales, and it is in complete agreement with solution (\ref{bfv2}) in Sect.~\ref{sNTPVs} and with~\cite{P1,P2}. Therefore, as expected, incorporating the right-hand side of (\ref{ltv''1}) into the calculation only changes the decaying mode of the solution.

\subsubsection{Quasi-Newtonian treatment}\label{sssQNT}
Expression (\ref{GRdotv}) is also the starting point of the quasi-Newtonian studies of linear peculiar velocities in cosmology~\cite{M,EvEM}. However, the agreement with the relativistic analysis stops at this point because of the strict constraints imposed on the perturbed spacetime by the quasi-Newtonian approximation. These constraints compromise the relativistic nature of the analysis and eventually lead to Newtonian equations and results (see Sect.~6.8.2 in~\cite{EMM} - pages 150 and 151 - for related `warning' comments). More specifically, the quasi-Newtonian approach starts with setting the vorticity to zero at the linear level. This has the benefit of simplifying the mathematics because it allows one to express the four-acceleration as the spatial gradient of a scalar potential ($\varphi$) so that
\begin{equation}
A_a= {\rm D}_a\varphi\,. \label{qN4A}
\end{equation}
There is a downside, however, because $\varphi$ is an ad hoc potential with no expression to describe it, while its linear evolution follows from the (non-uniquely determined) ansatz $\dot{\varphi}=-\Theta/3$~\cite{M}. In addition, the aforementioned ansatz requires setting the linear shear to zero as well. There is more to it, however. When it comes to the study of peculiar velocities, the most serious downside of the quasi-Newtonian treatment is that it inadvertently bypasses the gravitational contribution of the peculiar flux. This severely compromises the relativistic nature of the study and makes it identical to its purely Newtonian counterpart for all practical purposes. Indeed, combining (\ref{GRdotv}) and (\ref{qN4A}) the differential equation governing the linear evolution of the peculiar-velocity field takes the form~\cite{M,EvEM}
\begin{equation}
\ddot{v}_a+ 3H\dot{v}_a- H^2v_a= 0\,,  \label{qNddotv}
\end{equation}
and accepts the power-law solution
\begin{equation}
v= \mathcal{C}_1t^{1/3}+ \mathcal{C}_2t^{-4/3}= \mathcal{C}_3a^{1/2}+ \mathcal{C}_4a^{-2}\,,  \label{qNv}
\end{equation}
both of which are identical to their purely Newtonian analogues (compare to Eqs.~\ref{ltddotv2} and \ref{lNv2} in Sect.~\ref{sssNT}). Therefore, according to the Newtonian and the quasi-Newtonian studies, linear peculiar velocities grow no faster than $v\propto t^{1/3}\propto a^{1/2}$ during the Einstein-de Sitter era. However, this growth rate is not strong enough to explain the fast and deep bulk flows reported by several surveys over roughly the past fifteen years.

\subsubsection{Relativistic treatment}\label{sssRT}
The relativistic analysis of linear peculiar velocities imposes no constraints on the perturbed spacetime, and crucially, it also accounts for the gravitational input of the peculiar flux~\cite{TT,MiT,T}. The latter feeds into Einstein's equations and emerges into the equations monitoring a cosmological model endowed with a peculiar-velocity field. As a result, there is no longer any need to introduce an ad hoc scalar potential ($\varphi$) for the four-acceleration (see Eq.~\ref{qN4A}) or adopt the ansatz $\dot{\varphi}=-\Theta/3$ for its time evolution. Instead, the four-acceleration emerges naturally in the linear formulae governing the evolution of the density and the expansion gradients. These are respectively given by~\cite{T}
\begin{equation}
\dot{\Delta}_a= -\mathcal{Z}_a- 3aHA_a- a{\rm D}_a\vartheta  \label{GRlDela}
\end{equation}
and
\begin{equation}
\dot{\mathcal{Z}_a}= -2H\mathcal{Z}_a- {3\over2}\,H^2\Delta_a- {9\over2}\,aH^2A_a+ a{\rm D}_a{\rm D}^bA_b\,,  \label{GRlcZa}
\end{equation}
with $\Delta_a=(a/\rho){\rm D}_a\rho$ and $\mathcal{Z}_a=a{\rm D}_a\Theta$ representing perturbations in the matter density and the universal expansion, respectively.\footnote{Expressions (\ref{GRlDela}) and (\ref{GRlcZa}) follow by respectively linearising the spatial gradient of the energy conservation law (see (\ref{lcls}) in Sect.~\ref{ssKRPF}) and that of the Raychaudhuri equation~\cite{T}. Alternatively, one can obtain (\ref{GRlDela}) and (\ref{GRlcZa}) by linearising the non-linear formulae (2.3.1) and (2.3.2) of~\cite{TCM}, or
Eqs.~(10.101) and (10.102) of~\cite{EMM} while taking into account the linear relation (\ref{lcls}b) and that $q_a=\rho v_a$ in our case.} Also, $\vartheta={\rm D}^av_a$ is the spatial divergence of the peculiar velocity. Expressions (\ref{GRlDela}) and (\ref{GRlcZa}) provide the relativistic set of differential equations that govern the linear evolution of peculiar velocities in a perturbed Einstein-de Sitter universe. It is therefore clear that the relativistic analysis naturally couples the driving force of the peculiar-velocity field, namely the four-acceleration, to both the density and the expansion gradients.

It is important to note at this point that, had the gradient of the energy-conservation law (\ref{lcls}a) been involved in~\cite{M}, that study would also have arrived at the relativistic expression (\ref{GRlDela}) for the 4-acceleration, instead of the Newtonian-like (\ref{qN4A}) - see also footnote~5 here. This omission reduced the quasi-Newtonian analysis to purely Newtonian for all practical purposes.

By differentiating (\ref{GRlDela}) with respect to time, substituting the right-hand side of (\ref{GRlcZa}) in the resulting expression, and then using Eq.~(\ref{GRdotv}), we arrived at~\cite{T}
\begin{equation}
\ddot{v}_a+ H\dot{v}_a- {3\over2}\,H^2v_a= {1\over3aH}\left(\ddot{\Delta}_a+2H\dot{\Delta}_a -{3\over2}\,H^2\Delta_a\right)\,,  \label{GRlddotv1}
\end{equation}
which is the relativistic formula governing the evolution of linear peculiar velocities in an Einstein-de Sitter cosmology. We note that in deriving the above equation, we also used the background relation $\dot{H}=-3H^2/2$ together with the linear commutation laws $\partial_t({\rm D}^av_a)={\rm D}^a\dot{v}_a-H{\rm D}^av_a$ and $\partial_t({\rm D}_a\vartheta)={\rm D}_a\dot{\vartheta}-H{\rm D}_a\vartheta$.

The expression (\ref{GRlddotv1}) is a non-homogeneous differential equation, and its homogeneous component of reads as\footnote{The terms homogeneous andnon-homogeneous refer to the nature of the differential equation and not to the homogeneity (or lack thereof) of the host spacetime. We remind the reader that the latter is both inhomogeneous and anisotropic at the linear level.}
\begin{equation}
\ddot{v}_{\alpha}+ H\dot{v}_{\alpha}- {3\over2}\,H^2v_{\alpha}= 0\, \label{GRlddotv2}
\end{equation}
and accepts the solution
\begin{equation}
v= \mathcal{C}_1t+ \mathcal{C}_2t^{-2/3}= \mathcal{C}_3a^{3/2}+ \mathcal{C}_4a^{-1}\, \label{GRlv1}
\end{equation}
on all scales~\cite{T}. Therefore the relativistic analysis has led to the considerably stronger growth rate of $v\propto t$ for linear peculiar velocities. Moreover, according to the theory of differential equations, solution (\ref{GRlv1}) provides the minimum growth rate of the $v$ field. Indeed, a fairly well known theorem states that the full solution of a non-homogeneous differential equation forms from the general solution of its homogeneous part--in our case from (\ref{GRlv1})--and from a partial solution of the full equation. Therefore, solving (\ref{GRlddotv1}) in full will make a physical difference only if its partial solution grows faster than the fastest growing mode of the homogeneous solution. Put another way, the power-law $v\propto t$ is the minimum growth rate of the linear peculiar-velocity field. Indeed, by following an approach that accounts for some (though not all) of the effects of the density gradients seen on the right-hand side of (\ref{GRlddotv1}), the studies of~\cite{TT} and \cite{MiT} led to the growth rate of $v\propto t^{4/3}$ for the linear $v$ field. Overall, the relativistic analysis leads to considerably stronger growth rates for peculiar velocities than the Newtonian and quasi-Newtonian treatments. This means that general relativity supports faster and deeper residual bulk flows than is generally anticipated, perhaps the bulk flows such as those reported in the surveys of~\cite{WFH}-\cite{WHD}.

\subsubsection{The role of the four-acceleration}\label{sssR4-A}
A complementary, less technical, but perhaps more physically intuitive way of demonstrating and explaining the faster growth rates of the relativistic analysis is by looking closer at the role of the four-acceleration. According to linear theory, the latter is the driving force of the $v$ field (see Eq.~\ref{GRdotv} in Sect.~\ref{ssNQ-NRTs}). Therefore, the stronger the four-acceleration, the faster the peculiar velocity.

Following Eqs.~(\ref{GRlDela}) and (\ref{GRlcZa}), the input of the peculiar flux to the relativistic gravitational field also couples the peculiar four-acceleration to the density and the expansion gradients. If we ignore for the moment the effects of the aforementioned gradients and keep only that of the peculiar flux, then on using (\ref{GRdotv}), expressions (\ref{GRlDela}) and (\ref{GRlcZa}) reduce to
\begin{equation}
A_a= -{1\over3H}\,{\rm D}_a\vartheta \hspace{5mm} {\rm and} \hspace{5mm} 9H^2A_a= -2{\rm D}_a\dot{\vartheta} -4H{\rm D}_a\vartheta\,,  \label{Aa3}
\end{equation}
both of which combine to the linear relation
\begin{equation}
{\rm D}_a\dot{\vartheta}= -{1\over2}\,H{\rm D}_a\vartheta\,.  \label{laux}
\end{equation}
By taking the time derivative of (\ref{Aa3}a), employing Eq.~(\ref{laux}), and using the auxiliary background and linear relations given in Sect.~\ref{sssRT} previously, one can easily show that $\dot{A}_a=0$. Accordingly, when only the flux-effects are accounted for, linear peculiar velocities are driven by a time-invariant four-acceleration. In such cases, the differential equation (\ref{GRdotv}) accepts the linearly growing solution
\begin{equation}
v= \mathcal{C}_1t+ \mathcal{C}_2t^{-2/3}\,  \label{GRfv}
\end{equation}
on all scales and in full agreement with solution (\ref{GRlv1}) obtained in Sect.~\ref{sssRT}.

Keeping the density and the expansion gradients on the right-hand side of (\ref{GRlDela}) and taking its time derivative, we arrived at the following propagation formula for the linear peculiar four-acceleration~\cite{TT}:
\begin{equation}
\dot{A}_a- {1\over2}\,HA_a= -{1\over3H}\,{\rm D}_a\dot{\vartheta}- {1\over3aH}\left(\ddot{\Delta}_a+\dot{\mathcal{Z}}_a\right)\,.  \label{dotAa}
\end{equation}
Solving the homogeneous left-hand side of the above differential equation gives $A_a\propto t^{1/3}$ (we recall that $H=2/3t$ in the background). Finally, after substituting $A_a\propto t^{1/3}$ into the right-hand side of Eq.~(\ref{GRdotv}), we found that
\begin{equation}
v= \mathcal{C}_3t^{4/3}+ \mathcal{C}_2t^{-2/3}\,  \label{GRgv}
\end{equation}
in agreement with~\cite{TT} (see also~\cite{MaT,MiT}). Consequently, when some of the gradient effects are also accounted for, the driving force of the peculiar-velocity could grow in time (as $A_a\propto t^{1/3}$). As a result, the associated $v$ field could grow faster than linearly, namely as $v\propto t^{4/3}$ - a possibility worthy of further investigation.

Contrary to its relativistic counterparts, the quasi-Newtonian four-acceleration decays in time as $A_a\propto t^{-2/3}$~\cite{M}. Substituting this result into (\ref{GRdotv}), the latter leads to a linear peculiar-velocity field that grows as $v\propto t^{1/3}$ only. It goes without saying that the purely Newtonian studies lead to the same exact results (e.g.~see Eq.~\ref{Npaccel2} in Sect.~\ref{sNTPVs}). The fact that in the Newtonian and quasi-Newtonian studies, the linear peculiar velocities are driven by a decaying force, whereas the relativistic driving agent is constant or even increasing in time, explains both the rather mediocre $v\propto t^{1/3}$ growth rate of the former treatments and the substantially stronger ($v\propto t$, or even $v\propto t^{4/3}$) rates of the latter.

\subsubsection{The peculiar vorticity and shear}\label{sssPVS}
A generic peculiar-velocity field is expected to have non-zero vorticity and shear, and they are respectively defined as the antisymmetric and the symmetric traceless components of the $v$ field. In other words, $\varpi_{ab}={\rm D}_{[b}v_{a]}$ is the peculiar vorticity, and $\varsigma_{ab}={\rm D}_{\langle b}v_{a\rangle}$ is the peculiar shear. These kinematic variables do not necessarily evolve as the peculiar velocity itself. Indeed, by taking the spatial gradient of (\ref{GRlddotv2}) and using the linear commutation laws ${\rm D}_b\dot{v}_a=\partial_t({\rm D}_bv_a)+H{\rm D}_bv_a$ and ${\rm D}_b\ddot{v}_a=\partial_{tt}({\rm D}_bv_a)+2H\partial_{t}({\rm D}_bv_a)-(H^2/2){\rm D}_bv_a$, we arrive at the following differential equation
\begin{equation}
\ddot{\varpi}_{ab}+ 3H\dot{\varpi}_{ab}- H^2\varpi_{ab}= 0\,  \label{lddotvpi}
\end{equation}
for the linear evolution of the peculiar vorticity. Clearly, an exactly analogous differential formula also monitors the linear peculiar shear ($\varsigma_{ab}$). Recalling that $H=2/3t$ during the Einstein-de Sitter epoch, when the above equation is solved, it gives
\begin{equation}
\varpi= \mathcal{C}_1t^{1/3}+ \mathcal{C}_2t^{-4/3}= \mathcal{C}_3a^{1/2}+ \mathcal{C}_4a^{-2}\,,  \label{lvpi}
\end{equation}
with the same solution governing the evolution of the peculiar shear ($\varsigma$) as well. Therefore, both $\varpi$ and $\varsigma$ grow considerably slower than the peculiar-velocity field itself (we recall that $v\propto t$ -- see solution (\ref{GRlv1}) in Sect.~\ref{sssRT} before). This result is in agreement with the earlier relativistic analysis of \cite{TT}.

\subsubsection{Impact of the late-time
acceleration}\label{sssILTA}
So far our analysis has been confined to the Einstein-de Sitter epoch of the universe. Within the framework of the $\Lambda$CDM scenario, the earlier dust-dominated era is followed by a recent phase of accelerated expansion, where the energy density of the cosmos is typically dominated by a cosmological constant or by dark energy. Therefore, our results apply to the period between matter--radiation equality and the onset of cosmic acceleration. Once the latter epoch has started, the evolution of the peculiar-velocity field is expected to change drastically. More specifically, cosmic acceleration is expected to suppress the growth of peculiar velocities, in the same way it suppresses essentially all types of perturbations, such as those in the matter density.

Analytical relativistic studies that directly address and quantify the impact of a late $\Lambda$CDM phase on large-scale peculiar velocities would be very useful. At present, there is support for the aforementioned suppressing effect of the universal acceleration, but it is indirect. In particular, according to the Newtonian numerical work of~\cite{Detal}, dark energy inhibits the growth of peculiar velocities compared to the Einstein-de Sitter epoch. Studies looking into the linear evolution of peculiar velocities during a phase of de Sitter inflation have also shown exponential decay for the $v$ field~\cite{MaT}. Nevertheless, even if peculiar velocities are suppressed during the late $\Lambda$-dominated phase, at least some of the strong growth of the preceding Einstein-de Sitter period should survive through to the present.

On these grounds, one should expect to measure bulk velocities that are faster than anticipated by the Newtonian studies at higher redshifts but to also see a decrease in their mean value at lower redshifts (see~\cite{T,TPA} for further discussion). We note that a peculiar-velocity profile with increased values at higher redshifts (i.e.~at earlier times) and a late-time suppression has been reported in the survey of~\cite{CMSS} and (more pronounced) in those of~\cite{Wetal,WF}. Put another way, our work appears to provide theoretical support to the aforementioned observational studies.

\section{An effective gravitational potential}\label{sEGP}
As we have already mentioned, it is possible to recover the relativistic solution through Newtonian analysis, although the mapping is neither unique nor physically equivalent in a strict sense. To do so, one needs to introduce a specific ansatz for the evolution of the gravitational potential (or, equivalently, for the integration constants that appear when Poisson's equation is used in reverse). More specifically, in line with the covariant Newtonian study (see Sect.~\ref{sssNT}), the divergence of Eq.~(\ref{graddotPhi}) gives
\begin{equation}
\partial^2\dot{\phi}= {1\over2}\,H\partial^2\phi+ {3\over2}\,H\partial^{\alpha}\dot{v}_{\alpha}\,,  \label{efphi1}
\end{equation}
which upon integrating once in space yields
\begin{equation}
\partial_{\alpha}\dot{\phi}= {1\over2}\,H\partial_\alpha\phi+ {3\over2}\,H\dot{v}_{\alpha}+ V_{\alpha}\,.  \label{efphi2}
\end{equation}
Here $V_{\alpha}$ could (in principle) have both a temporal and a spatial dependence, and it encodes the solenoidal (divergence-free) integration mode that is left undetermined by the divergence in Eq.~(\ref{efphi1}) and is fixed only after specifying boundary conditions and/or a coordinate choice. When substituting the above equation into the right-hand side of (\ref{ltv''1}), one arrives at the following (non-homogeneous) differential equation
\begin{equation}
\ddot{v}_{\alpha}+ 3H\dot{v}_{\alpha}- H^2v_{\alpha}= -V_{\alpha}\,  \label{efphi3}
\end{equation}
for the linear evolution of the peculiar-velocity field. Expression (\ref{efphi3}) differs from Eq.~(\ref{ltddotv2}) only by its non-zero right-hand side, the presence of which allows one to change the standard Newtonian evolution of the linear peculiar velocities. By allowing $V_\alpha\neq0$, we effectively parameterise the freedom that the time evolution of $\nabla \phi$ is not fixed by the Poisson equation alone. For instance, by choosing $V_{\alpha}=-(5/2)H\dot{v}_{\alpha}-H^2v_{\alpha}$, one can recover the relativistic growth rate of $v\propto t$ obtained previously in Sect.~\ref{sssRT} (see Eqs.~(\ref{GRlddotv2}) and (\ref{GRlv1}) there). It goes without saying, however, that our choice of $V_{\alpha}$ is not uniquely determined, and therefore this procedure should be viewed as illustrative mapping rather than as a physical equivalence between Newtonian gravity and general relativity.

It is worth pointing out that a vector function analogous to $V_{\alpha}$ was obtained in~\cite{P2} also after involving and integrating Poisson's formula. There, the aforementioned vector was set to zero by an appropriate choice of coordinates or boundary conditions. This freedom is reminiscent of the fact that the relativistic peculiar flux is also a vector that can be set to zero (albeit only locally) by moving to the energy (or Landau-Lifshitz) frame. Moreover, even in the energy frame, the peculiar energy flux does not disappear but re-emerges `disguised' as a `peculiar particle flux'. We remind the reader that in relativity and depending on the specifics of the problem in hand, one may adopt the energy frame where the energy flux vanishes but there is non-zero particle flux. Alternatively, one may choose to use the particle (or Eckart) frame when the situation is reversed (e.g.~see~\cite{TCM,EMM}).

\section{Discussion}\label{sD}
Large-scale peculiar motions, commonly referred to as bulk flows, are commonplace in our Universe and have been repeatedly verified by many surveys (e.g.~see~\cite{ND}-\cite{WHD} for a representative, though incomplete, list). Although the aforementioned flows typically encompass domains of up to several hundred megaparsecs, we believe that they started as weak peculiar-velocity perturbations around the time of recombination that grew larger and faster by structure formation and by the increasing non-uniformity of the post-recombination Universe. Nevertheless, the evolution and the implications of these bulk peculiar motions are still largely unknown. Moreover, during the roughly past fifteen years the bulk-flow puzzle has become even more perplexing, after an increasing number of surveys have reported sizes and speeds well in excess of those allowed by the concordance cosmological model, namely, by the $\Lambda$CDM paradigm. In fact, if one is allowed to take into account the controversial dark flows, with sizes of the order of 1000~Mpc and speeds close to 1000~km/sec (see~\cite{Ketal,KA-BE} for example), the situation becomes rather untenable.

Significantly, the $\Lambda$CDM constraints on the speed and the depth of the observed bulk peculiar flows are based entirely on Newtonian theory, and there is a serious difference between Newtonian gravity and general relativity when it comes to the study of peculiar motions, a difference which stems from the fundamentally different way the two theories treat the gravitational field and its sources. To begin with, we recall that bulk flows are nothing but matter in motion and that moving matter means a non-zero energy flux. In Newton's theory only the density of the matter contributes to the gravitational field, via Poisson's equation, while any energy fluxes that may exist have no gravitational input. In relativity, on the other hand, the pressure (both isotropic and viscous), as well as the fluxes also contribute to the energy momentum tensor and therefore to the (relativistic) gravitational field. It is the input of the peculiar flux that has been bypassed in essentially all the mainstream studies so far. Thus, accounting for the gravitational contribution (or not) is the key difference that separates the relativistic treatment from the Newtonian and quasi-Newtonian treatments of cosmological peculiar motions.

By not incorporating the gravitational input of the peculiar flux, Newtonian and quasi-Newtonian analyses have both led to the same mediocre ($v\propto t^{1/3}$) growth rate for the linear peculiar-velocity field~\cite{P1}-\cite{EvEM}. Unfortunately, the $v\propto t^{1/3}$ rate is too slow to explain the reported fast bulk flows. However, had the gradient of the energy-conservation law (\ref{lcls}a) been included in~\cite{M,EvEM}, these quasi-Newtonian studies would also have arrived at the relativistic expression (\ref{GRlDela}) for the 4-acceleration, instead of the Newtonian (\ref{qN4A}) - see Sect.~\ref{sssQNT} and \ref{sssRT}. In that case, the quasi-Newtonian and the relativistic treatments would have arrived at the same results and any subsequent comparison between them would have been entirely unnecessary.

On the other hand, by including the peculiar-flux contribution to the gravitational field alone, the relativistic analysis has led to the considerably faster growth rate of $v\propto t$ (see~\cite{T}) and also Sect.~\ref{sssRT} here). Moreover, the standard theory of differential equations guarantees that $v\propto t$ is the minimum growth supported by the relativistic studies. To a large extent, the latter conclusion was already foreseen in the studies of~\cite{TT}-\cite{MiT}, which arrived at the faster linear growth rate of $v\propto t^{4/3}$, after accounting for some of the inhomogeneity effects as well.

A complementary explanation for the aforementioned relativistic results involves the acceleration and the four-acceleration, which are the driving forces of the linear peculiar-velocity field in all three approaches. A closer inspection (see Sect.~\ref{sssR4-A}) revealed that the driving agent decays as $\propto t^{-2/3}$ in both the Newtonian and the quasi-Newtonian studies. In contrast, the relativistic four-acceleration is constant in time (at a minimum; see~\cite{T} and Sect.~\ref{sssRT} here), or it could potentially increase as $t^{1/3}$ throughout the Einstein-de Sitter phase of the Universe (see~\cite{TT} and~\cite{MiT}). This drastic difference explains why general relativity supports significantly faster peculiar-velocity fields. We also note that within the framework of the standard $\Lambda$CDM model, cosmic acceleration starts late into the Einstein-de Sitter phase of the Universe, and some (at least) of the increased velocity growth taking place during the Einstein-de Sitter epoch should survive through the subsequent accelerated phase and have consequently led to faster and deeper residual bulk flows today, such as those reported in~\cite{WFH}-\cite{WHD}. Moreover, all of this happens within the standard $\Lambda$CDM model and without the need for any new free parameters.

The exact analytical results of the linear analysis presented here as well as in the earlier works of~\cite{TT}-\cite{MiT}, and~\cite{T} can also provide a useful testing ground for the emerging numerical studies simulating the post-recombination universe. The fact that all the aforementioned theoretical treatments are gauge invariant is an additional advantage since their results are free of any gauge-related ambiguities.\footnote{The peculiar-velocity field ($v$) vanishes by default in the FRW background, which makes it a gauge-invariant linear perturbation~\cite{SW}.} We recall that (to the best of our knowledge) the numerical simulations rely on particular gauge choices when addressing the problems in hand.

Up to now, most of the simulations that employ relativistic techniques do not consider the evolution of peculiar velocities, with a couple of possible exceptions that looked at the effects of the peculiar vorticity (e.g.~\cite{B-Hetal}). The vorticity was found to have a negligible impact, which was to be expected since it grows considerably slower that the peculiar velocity itself (even in the relativistic analysis; see~\cite{TT} and also Sect.~\ref{sssPVS} here). Moreover, the simulations typically operate within the $\Lambda$CDM paradigm (e.g.~see~\cite{ACL} for a recent review), so the numerical codes are adapted to the recent accelerated phase of the Universe, where the peculiar-velocity field is suppressed (e.g.~see~\cite{Detal}), instead of the preceding Einstein-de Sitter epoch, where peculiar velocities are strongly enhanced. As a result, so far there is no common ground to facilitate direct comparison between analytical and the numerical work on this topic. Nevertheless, the relativistic simulations are still under development, and there is room for testing and improvement, especially in view of repeated surveys reporting bulk flows in excess of the (Newtonian-based) $\Lambda$CDM limits. Hopefully, the results of the analytical relativistic studies will provide the motivation for extending and adapting the numerical codes to address the ongoing bulk-flow puzzle.

Observationally, the time evolution of the peculiar velocities can be mapped into a redshift evolution at low $z$, enabling a comparison between our theoretical predictions and the standard Newtonian expectation using peculiar-velocity surveys such as CosmicFlows-4. However, peculiar-velocity measurements are subject to a well-known degeneracy between the assumed background cosmology and local kinematic contributions. This degeneracy can be partially mitigated through independent distance indicators, such as Type Ia supernovae or Tully-Fisher relations, which provide redshift-independent distance estimates. Nevertheless, the use of such indicators introduces additional uncertainties that grow with redshift, thereby impacting the reconstruction of the peculiar-velocity field and limiting the robustness of the physical inferences. A detailed implementation of unbiased and statistically robust estimators — such as the minimum-variance reconstruction techniques applied to CosmicFlows-4~\cite
{Wetal,WF} — is beyond the scope of the present work and will be addressed in future studies.

In summary, general relativity can provide a simple and physically motivated theoretical solution to the ongoing bulk-flow puzzle by simply taking into account the (purely general relativistic) gravitational contribution of the peculiar flux. This can in turn relax the current (purely Newtonian) peculiar-velocity expectations of the $\Lambda$CDM model and, in so doing, allow the latter to naturally accommodate bulk flows faster and deeper than anticipated, such as those reported in~\cite{WFH}-\cite{WHD}. The perceived discrepancy between the current cosmological model and the above bulk-flow reports thus may not reflect a generic problem in the $\Lambda$CDM paradigm, but instead indicate the use of an inappropriate gravitational theory in our studies of cosmological peculiar flows.\\

\textbf{Acknowledgements:} CGT acknowledges support from the Hellenic Foundation for Research and Innovation (H.F.R.I.), under the ``First Call for H.F.R.I. Research Projects to support Faculty members and Researchers and the procurement of high-cost research equipment Grant'' (Project Number: 789).

\end{document}